Increase in cratering efficiency with target curvature in strength-controlled craters

A. I. Suzuki [a,*], C. Okamoto [b], K. Kurosawa [c], T. Kadono [d], S. Hasegawa [a] and T. Hirai [a]

[a] Institute of Space and Astronautical Science, Japan Aerospace Exploration Agency, 3-1-1 Yoshinodai, Chuo, Sagamihara 252-5210, Japan

[b] Graduate School of Science, Kobe University, 1-1 Rokkodai, Nada, Kobe 657-8501, Japan

[c] Planetary Exploration Research Center, Chiba Institute of Technology, 2-17-1 Tsudanuma, Narashino, Chiba 275-0016, Japan

[d] School of Medicine, University of Occupational and Environmental Health, 1-1 Iseigaoka, Yahata, Kitakyusyu 807-8555, Japan

* Corresponding author.

E-mail address: suzuki.ayako@isas.jaxa.jp (A. I. Suzuki)

Institute of Space and Astronautical Science, Japan Aerospace Exploration Agency, 3-1-1 Yoshinodai, Chuo, Sagamihara 252-5210, Japan

TEL: +81-50-3362-5945

FAX: +81-42-759-8454

Keywords: Impact processes; Cratering; Asteroids; surface

ABSTRACT

Impact-cratering processes on small bodies are thought to be mainly controlled by the local material strength because of their low surface gravity, and craters that are as large as the parent bodies should be affected by the target curvature. Although cratering processes on planar surfaces in the strength-controlled regime have been studied extensively, the mechanism by which target curvature affects the cratering processes remains unclear. Herein, we report on a series of impact experiments that used spherical targets with various diameters. The resultant craters consisted of a deep circular pit and an irregular-shaped spall region around the pit, which is consistent with the features reported in a number of previous cratering experiments on planar surfaces. However, the volume and radius of the craters increased with the normalized curvature. The results indicate that the increase in the spall-region volume and radius mainly contributes to the increase in the whole crater volume and radius, although the volume, depth, and radius of pits remain constant with curvature. The results of our model indicate that the geometric effect due to curvature (i.e., whereby the distance from the equivalent center to the target free surface is shorter for higher curvature values) contributes to increases in the cratering efficiency. Our results suggest that the impactors that produce the largest craters (basins) on some asteroids are thus smaller than what is estimated by current scaling laws, which do not take into account the curvature effects.

1. Introduction

Recent planetary explorations have revealed detailed surface morphologies of small bodies, which are covered by craters with diameters ranging from sub-micrometer sizes (e.g., Nakamura et al., 2012) to diameters exceeding the radius of the parent body (e.g., Leliwa-Kopystyński et al., 2008). Studies of impact craters on small bodies allow us to elucidate their collisional history and provide basic knowledge of the process of planetary accretion.

Impact-cratering processes on small bodies are expected to differ significantly from those on terrestrial planets mainly for the following two reasons. The first is related to their relatively low surface gravity; in other words, the local material strength rather than gravity controls the crater size for small bodies. All craters on asteroids smaller than a few kilometers in diameter may have been produced in the strength-controlled regime (Jutzi et al., 2015). The second reason involves the effect of target curvature. When the radius of a crater exceeds that of the parent body, the effects of curvature on the impact process cannot be neglected (e.g., Cintala et al., 1978; Fujiwara et al., 1993). Thus, understanding cratering processes on curved surfaces in the strength-controlled regime is essential for investigating the history of the collisional environment of small bodies.

Although cratering processes on planar surfaces in the strength-controlled regime have been studied extensively (e.g., Shoemaker et al., 1963; Hörz, 1969; Gault, 1973; Lange et al., 1984; Polanskey and Ahrens, 1990; Baldwin et al., 2007; Milner et al., 2008; Suzuki et al., 2012; Dufresne et al., 2013; Poelchau et al., 2014), the number of studies on the effects of curvature on the cratering processes is relatively small. Fujiwara (1991) made theoretical predictions of the radii of craters on curved surfaces. In addition, Fujiwara et al. (1993, 2014) produced distinctive impact craters on mainly cylindrical targets with a wide range of radii in a laboratory and presented the empirical relations between crater diameter, depth, mass, and target curvature. Walker et al. (2013) had an aluminum sphere impact into granite spheres of 1-m diameter at 2 km/s in order to examine the scale size effect of momentum enhancement in the momentum transfer in impacts.

In this work, we report on a series of impact experiments that were conducted by using spherical targets of various diameters to investigate how curvature affects the cratering efficiency. The three-dimensional topography of craters produced on the spherical surfaces were measured at a resolution of 0.2 mm/pixel, which is 16 times smaller than the projectile diameter, and thus, we were able to investigate the crater dimensions as a function of the ratio of the projectile size to target size. Finally, we developed a simple semi-analytical model that describes how target geometry affects the crater radius.

2. Impact experiments

Impact experiments were done by using a two-stage light-gas gun called the Horizontal Accelerator for Science and Engineering (HASE), which is located at the Institute of Space and Astronautical Science (ISAS), Japan Aerospace Exploration Agency (JAXA) (Hasegawa, 2015). Figure 1 shows a schematic cross section of the experimental setup.

To control the target curvature, gypsum targets ($CaSO_4 \cdot 2H_2O$) of various sizes and shapes were prepared as follows. Tap water and commercial $CaSO_4 \cdot 0.5H_2O$ powder were mixed at a mass ratio of 1:1.35, and the slurry was poured into foamed-polystyrene molds.

Six molds of various sizes and shapes were used, including cubes with sides of 9 and 15 cm, spheres with diameters of 7.8 and 10.9 cm, and hemispheres with diameters of 17.0 and 24.8 cm. The targets were removed from the molds after several hours and dried at room temperature with no artificial air currents for at least three days. Hereafter, the cubic and spherical targets are referred to as C9, C15, S8, S11, S17, and S25, respectively. The normalized curvature $\chi$ of the targets is defined as the ratio of the projectile diameter to target diameter. The normalized curvatures of C9 and C15 were zero, and those of S8, S11, S17, and S25 were 0.013, 0.019, 0.029, and 0.041, respectively.

To determine the bulk density, porosity, and tensile strength of the targets, disk-shaped specimens of $CaSO_4 \cdot 2H_2O$ were also prepared by using the same method. These specimens were 25 mm in diameter and 10 mm thick. The mean bulk density of 17 specimens was $1.08 \pm 0.01$ g/cm$^3$. The porosity was calculated to be $53.1\% \pm 0.5\%$ based on a grain density of gypsum of $2.304 \pm 0.002$ g/cm$^3$, as measured by an AccuPyc II 1340 gas displacement pycnometer (Micrometrics). The indirect diametrical compression test (so-called Brazilian disc test) was adopted to measure the tensile strength of the specimens (e.g., Wong and Jong, 2014). The specimens were compressed in a diametrical direction at a loading rate of 1 mm/min with a uniaxial-compressive testing machine at ISAS/JAXA. Following increases in the applied force, the specimens were split into two major pieces by the internal tensile force acting perpendicular to the loading direction. The indirect tensile strength is $\sigma_t = 2F/(\pi d h)$, where $F$ is the applied force at the split, and $d$ and $h$ are the diameter and thickness of the specimens, respectively. The average of the tensile strength for 10 specimens was $2.3 \pm 0.2$ MPa.

The targets were placed on a mechanical jack with a pedestal, and this equipment was set in a polystyrene box with an acrylic window on the side for observations and a small rectangular hole at the front through which the projectile passed. In-situ images were acquired with a high-speed digital video camera (Shimadzu, HPV-X or HPV-X2) through the window at a frame rate of 50 µs/frame. The polystyrene box was placed in a vacuum chamber, and the chamber was evacuated to 2–12 Pa prior to each shot. A spherical nylon projectile with a diameter of 3.2 mm and a weight of 0.019 g was accelerated horizontally to about 3.4 km/s by using a split-type nylon sabot (Kawai et al., 2010). Thus, all shots have almost identical kinetic energy. Table 1 summarizes the projectile and target conditions for each shot. Because we performed head-on collisions in this study, the impact angles of all shots were assumed to be 90 degrees in the following analyses, as measured from the plane tangent to the impact point (see Section 3). The impact points on the spherical targets, however, were scattered slightly within a few centimeters in diameter. For reference, the actual impact angles were measured from the high-speed images and are listed in Table 1. The angles of most shots exceeded 70 degrees.

The target and ejected fragments were recovered from the box after each shot. To obtain the crater dimensions, the target surface, including the crater, was scanned by a high-resolution three-dimensional geometry-measurement system consisting of a semiconductor laser displacement sensor and a stage controlled by two horizontal linear actuators (COMS MAP-3D). This device measured the height from a given point every 0.2 mm, which enabled us to obtain nonintrusive high-resolution measurements of the surface topography. We determined the equations of target spheres by using the raw data from outside of the crater and reproduced the pre-impact topography of the target spheres. The volume of the crater was derived by integrating the topographic difference between the pre- and post-impact surfaces. The depth of the crater is defined as the depth of the deepest point from the pre-impact surface. The radius of the crater is defined as the radius of a circle of the same surface area as the crater.

3. Experimental results

This section summarizes the results obtained in the impact experiments. First, we briefly describe the appearance of the ejecta curtains observed by the high-speed digital video camera. Next, we describe the crater dimensions including the volume, depth, radius, and cross-sectional profiles, which were used to characterize the crater morphology.

Figure 2 shows snapshots of the impact on a spherical target (#2699). We measured the impact angle of each shot from these images. An ejecta curtain forms an inverted cone on the spherical target as it evolves on the planar surface, as reported in a number of previous studies (e.g., Oberbeck and Morrison, 1976). The ejecta curtain consists of finer grains, while centimeter-sized spall fragments appear during the later times of fragment ejection. This also can be observed during impacts on planar surfaces (e.g., Polanskey and Ahrens, 1990).

Figure 3 shows photographs of typical resultant craters with various curvatures (on the same length scale). Each crater consisted of a deep circular pit and an irregular-shaped spall region around the pit. The pits had a rugged surface with cracks every 2–3 mm, which can be attributed to stress-wave propagation during impact. Conversely, the surfaces of the spall region were smooth and seemed to be the faces of the fractures. The surfaces of the spall region also exhibited a stepped morphology, and some of the large ejecta fragments fit exactly with the steps. Although the characteristics of the craters were basically consistent with those of craters formed on a planar surface of brittle materials (Dufresne et al., 2013 and references therein), the results indicate that the spall region increases with target curvature.

Figure 4 shows the cross-sectional topography of typical craters with various curvatures. The boundary between the pit and the spall region, which was determined manually from the point where the topographic slope changes drastically, is marked with open circles. The topographic profiles of the pits with various curvatures overlap well, in contrast to the extension of the spall region with increasing curvature. This trend is particularly evident for $\chi \geq 0.02$.

Table 2 provides the crater dimensions. Figure 5 illustrates (a) the volumes of whole craters, pits, and spall regions, (b) the depth of craters, and (c) the radius of craters with the normalized curvature. The volume and radius of whole craters increased with the normalized curvature, whereas the volume, depth, and radius of pits remained constant. In other words, the increase in the spall-region volume and radius mainly contributed to increases in the crater volume and radius. The volume of the spall region exceeded that of the pit for $\chi \gtrsim 0.019$ in spite of the fact that the pit volume was dominant on the crater formed on the surface for $\chi \lesssim 0.013$.

Figure 6 shows the crater volume ($H_v$), depth ($H_b$), and radius ($H_r$) on curved targets normalized by those on plane targets as a function of the normalized curvature. We also plotted the data derived from craters formed on the side of cylindrical gypsum by a nylon projectile that impacted the target at 3–4 km/s (Fujiwara et al., 2014). Normalized crater volume ($H_v$) and normalized radius ($H_r$) indicate a positive correlation with $\chi$ and are consistent with those obtained with the cylindrical targets. Normalized depth ($H_b$) remains constant within the range of the curvature in this study ($\chi < 0.041$), although Fujiwara et al. (2014) shows based on the results of a wider range of curvatures, impact velocities, and various targets and projectiles that $H_b$ slightly increases with curvature.

Here, we also present the data in terms of the Pi-group scaling laws. Four dimensionless parameters are introduced according to the Pi-group scaling theory (e.g., Melosh, 1989; Holsapple, 1993):

$$\pi_D = D\left(\frac{\rho_t}{m_p}\right)^{1/3}, \quad \pi_V = \frac{V\rho_t}{m_p}, \quad \pi_Y = \frac{Y}{\rho_t v_i^2}, \quad \pi_4 = \frac{\rho_t}{\rho_p}, \tag{1}$$

where $V$ and $D$ are the volume and diameter of the crater or pit, $m_p$, $\rho_p$, and $v_i$ are the projectile mass, density, and velocity, and $\rho_t$ and $Y$ are the target density and a strength measure, respectively. Although the Pi-group scaling laws are usually constructed from the dimensions of the transient craters, the volume and diameter of the final craters including the spall region have also been frequently used because these data are easily measured (e.g., Gault, 1973; Lange et al., 1984; Baldwin et al., 2007; Milner et al., 2008; Suzuki et al., 2012). We used the final size of the crater and pit in the following discussion. The scaled diameter $\pi_D$ and scaled volume $\pi_V$ can be expressed as a function of the normalized strength $\pi_Y$ and the density ratio $\pi_4$ as follows (Holsapple, 1993; Housen and

Holsapple, 2011):

$$\pi_D \propto \pi_Y^{-\frac{\mu}{2}} \pi_4^{\frac{1-3\nu}{3}}, \quad \pi_V \propto \pi_Y^{-\frac{3\mu}{2}} \pi_4^{1-3\nu}, \tag{2}$$

where $\mu$ and $\nu$ are scaling exponents related to a single measure $C = R_p v_i^\mu \rho_p^\nu$ called the "coupling parameter" (Holsapple, 1993). $R_p$ is the projectile radius.

In Fig. 7, the scaled cratering efficiencies of diameter $\pi_D/\pi_4^{(1-3\nu)/3}$ and of volume $\pi_V/\pi_4^{(1-3\nu)}$ for the crater and pit are plotted against $\pi_Y$ along with those on other brittle targets reported by previous studies (Gault, 1973; Suzuki et al., 2012; Yasui et al., 2012; Fujiwara et al., 2014). The value $\nu = 0.4$ was chosen because it has been well constrained in both gravity- and strength-controlled regimes (Schmidt, 1980; Holsapple and Schmidt, 1982; Schultz and Gault, 1985; Housen and Holsapple, 2011). The Pi-scaling equations of ejected mass and crater diameter for igneous rocks were derived by Suzuki et al. (2012) from the dimensional power-laws of them obtained by Gault (1973): $\pi_D = 0.95 \, \pi_Y^{-0.370} \pi_4^{-0.167}$ and $\pi_V = 0.03 \, \pi_Y^{-1.133} \pi_4^{-0.5}$. The tensile strength was used to calculate $\pi_Y$ in this study because it was expected to be the most important in the generation of spalled fragments. Note that the appropriate strength for $\pi_Y$ is still under debate (e.g., Güldemeister et al., 2015). If we chose to use different types of strength, such as dynamic strength (Güldemeister et al., 2015), $\pi_Y$ might be a somewhat lower value. Scaled cratering efficiencies of diameter and volume obtained from the whole craters at $\chi = 0$ were lower than the two regression lines for igneous (Gault, 1973) and sedimentary rocks (Suzuki et al., 2012). The high porosity of our targets (53%) might have led to such low cratering efficiencies. Further analyses are beyond the scope of this study.

4. Discussion

In the previous section, we showed that the crater volume and radius increase with the target curvature and that the extent of the spall region mainly contributes to the extent of the crater. In this section, we use a simple semi-analytical model to address the mechanism that explains the extent of the spall region as a function of the target curvature. Although Melosh (1984) presented a model to describe the size of the spall fragments, the direct application of this model to our results would be difficult because the rise time of the stress wave, which is the most important parameter in the model, is unknown in highly porous gypsum targets. Here, we discuss an alternative semi-analytical model to estimate the increased extent of the spall region formed on spherical targets compared to those formed on plane targets. We measured the crater size on the flat surface under the same impact conditions. The behavior of the stress wave in the spherical targets is expected to be the same as that in the flat targets. Thus, the curvature effects on the cratering efficiency can be addressed with a simple geometric consideration by reference to the data at $\chi = 0$. Then, we describe some prospects for planetary applications.

4.1 Simple model describing the geometric effects

When an impact occurs, a shock wave expands in the target as a hemisphere centered on *the equivalent center* (EC), and the shock wave attenuates with distance from the EC. According to Croft (1982), the pressure at a distance $l$ from the EC can be written as $P(l) = P_0 (l/\alpha R_p)^{-n}$, where $P_0$ is the peak pressure in the isobaric core, $R_p$ is the projectile radius, $\alpha$ is a parameter indicating the radius of the isobaric core as $\alpha R_p$, and $n$ is the pressure attenuation rate. We assume that the depth of EC equals the radius of the isobaric core. Since the spall plates are ejected by tensional stress (e.g., Melosh, 1984), we focus on the component of the force normal to the target surface. On an imaginary sphere centered at the EC, $P(l)$ can be regarded as the force applied per unit area. For a planar surface, the stress $P_n$ to the tangential plane at the crater rim is written as $P_n = P_0\{[(\alpha R_p)^2 + R_{plane}^2]^{1/2}/\alpha R_p\}^{-n} \sin\eta$, where $\eta$ is the angle between the target surface and the line between the EC and the crater rim, and $R_{plane}$ is the crater radius on the planar surface (Fig. 8a). This equation can be rewritten by using $R_n = R_{plane}/R_p$:

$$\frac{P_n}{P_0} = \left\{1 + \left(\frac{R_n}{\alpha}\right)^2\right\}^{-\frac{n+1}{2}}. \tag{3}$$

Conversely, for a curved surface of radius $R_t$, the stress $P'_n$ to the tangential plane at the crater rim is written as $P'_n = P_0 (l'/\alpha R_p)^{-n} \sin\eta'$, where $l'$ and $\eta'$ are the distance between the EC and the crater rim and the angle between the target surface and the line from the EC to the rim, respectively. They are written as follows (Fig. 8b):

$$l'^2 = R_t^2 + (R_t - \alpha R_p)^2 - 2R_t(R_t - \alpha R_p)\cos\omega, \tag{4}$$

$$l' \sin\eta' = (R_t/\cos\omega - R_t + \alpha R_p)\sin(\pi/2 - \omega), \tag{5}$$

where $\omega = R_{cr}/R_t$, and $R_{cr}$ is the crater radius measured along the target surface. Thus, $P'_n/P_0$ becomes

$$\frac{P_n'}{P_0} = \left\{1 + \frac{2(1-\alpha\chi)}{(\alpha\chi)^2}(1 - \cos\omega)\right\}^{-\frac{n+1}{2}} \left\{1 + \frac{1-\alpha\chi}{\alpha\chi}(1 - \cos\omega)\right\}. \tag{6}$$

Setting $P_n = P_n'$, the ratio $H_r$ defined as $R_{cr}/R_{plane} = R_t\omega/R_{plane} = \omega/(\chi R_n)$ can be derived as a function of $\chi$ for various $\alpha$ and $n$, where we set $R_n = 6.34$ according to our experimental results (Table 2).

Several research groups directly measured the pressure decay exponent with various materials at different ranges (e.g., Dahl and Schultz 2001; Kato et al., 2001; Nakazawa et al., 2002; Shirai et al., 2008). We calculated model curves for $n$ = 1.0, 1.5, 2.0 and $\alpha$ = 0.5, 1.0, 1.5, 2.0, and plots of selected curves with the experimental data are shown in Fig. 8c. The results indicate that (1) the ratio $H_r$ increases with the curvature, which is consistent with the experimental data, and (2) the lower attenuation rate and shallower EC enhances the rate of increase of $H_r$. The experimental data almost fall into the gray area in Fig. 8c where $n$ = 1.5 and $\alpha$ = 1.0–1.5. These values are consistent with the experimental results under the experimental conditions as follows. Gault and Heitowit (1963) estimated that the variation in pressure with distance tends to $n$ = 3/2 at a relatively low shock pressure (< 10 GPa). Senshu et al. (2002) modeled the range of $\alpha$ as $1 < \alpha < 1.44$ when the target and impactor are of the same material and the impact energy is perfectly converted to internal energy.

The good agreement between the experimental results and model predictions with reasonable parameters $n$ and $\alpha$ strongly suggests that the effect of target geometry (i.e., the distance from the EC to the target free surface being shorter for higher curvature values) mainly contributes to increases of the crater volume and radius with target curvature.

4.2 Implications for planetary craters

All the craters observed in this study had distinct spall regions, while craters with clear spalls have not been observed yet on planetary surfaces. Gault (1973), however, have reported that spall-like plates heaved upward and then settled down around artificial craters formed by TNT (trinitrotoluene) detonations on Earth. It is probable that spall plates are ejected from craters under a small gravitational field. Recently, spall craters have been considered as a third regime of cratering in addition to strength- and gravity-dominated regimes (Holsapple and Housen, 2013; Jutzi et al., 2015). Holsapple and Housen (2013) estimated that all the craters smaller than about 1 km are spall craters on a small body such as (433) Eros (16 km in diameter). Thus, spall-bearing craters on small bodies might be found by future planetary explorations. We believe that the semi-analytical model described here could become a useful tool to address the origins of such craters. It should be noted that the difference in sizes between targets used in a laboratory and natural planetary bodies is expected to be important because lower strain rate, longer shock-pressure-pulse duration, and the Weibull effect are expected to reduce the target strength at the planetary-scale impact event (e.g., Schultz and Gault, 1990; Housen and Holsapple, 1999; Poelchau et al., 2014). Such combination analysis of curvature and size effects is beyond the scope of this study.

In Fig. 9, we plot the experimental results on the diagram with parameters measurable on craters in the field; $H_{ch}$ is the ratio of the crater radius $R_{chord}$ measured as a chord on curved surfaces to those on planar surfaces. Another expression of curvature is $\varepsilon$, which is the ratio of the crater diameter $D_{chord}$ as a chord to the target radius. This value is commonly used to characterize large craters (basins) on small bodies (e.g., Leliwa-Kopystyński et al., 2008; Burchell and Leliwa-Kopystyński, 2010). Note that $D_{chord} = 2R_{chord} = 2R_t \sin(R_{cr}/R_t)$. Even in this figure, $H_{ch}$ increases with the curvature $\varepsilon$.

Our results reveal that the higher curvature causes the extension of the spall area, which results in the increase of the crater diameter. In other words, the impactor that produces a spall-dominant crater on a curved surface is expected to be smaller than that estimated by current scaling laws because the curvature effects are not taken into account in the scaling laws. Because the crater radii on curved surfaces at $\varepsilon = 0.9$ are, according to the experiment results, 1.7 times larger than those on planar surfaces, the estimated impactor mass can be a factor of 5 (i.e., $(1.7)^{-3} = 0.2$) smaller than that estimated by the scaling laws without the curvature effects. For example, $\varepsilon$ for some large craters (basins) on asteroids is roughly estimated to be ~0.86 (for a ~0.6 km crater on 243 Ida I Dactyl; Leliwa-Kopystyński et al., 2008) and ~0.70–0.84 (for a ~0.8 km crater on 4179 Toutatis; Hudson et al., 2003; Huang et al., 2013).

5. Summary

To investigate how target curvature affects cratering efficiency in strength-controlled craters, we performed a series of cratering experiments with spherical targets of various diameters and with impact velocities of 3–4 km/s. The resultant craters consisted of a deep circular pit and an irregular-shaped spall region around the pit, which is consistent with the results reported in a number of previous cratering experiments on planar surfaces. However, the results indicate that the volume and radius of whole craters increase with increases in the normalized curvature and that the extension of the spall region is responsible for the increase in the whole-crater volume and radius. Based on a simple model, we conclude that the effect of target geometry (i.e., the distance from the EC to the target free surface being shorter in the case of higher curvature values) represents the main contribution to the increase in crater volume and radius. Thus, the change in diameter due to the target curvature must be considered to accurately determine the size of impactors that produce large craters on small asteroids.

Acknowledgments

The series of experiments were supported by the Hypervelocity Impact Facility (the Space Plasma Laboratory) at ISAS/JAXA．We appreciate two anonymous referees for the constructive review that helped greatly improve the manuscript, and Francis Nimmo for helpful suggestions as an editor. AIS is supported by JSPS KAKENHI Grant Number JP24740307. KK is supported by JSPS KAKENHI Grant numbers JP17H02990, JP17K18812, JP15H01067, JP26610184 and JP25871212. We appreciate for useful discussion in the workshop on planetary impacts held at the Institute of Low Temperature Science Hokkaido University and Kobe University. We thank Nobuyuki Tokuoka (Shimadzu Corp.) for his excellent support as regards the in-situ observations by a high-speed digital video camera. We also thank Noboru Asakawa (Shimadzu Corp.) for his dedicated support to measure the grain density of gypsum targets.

Leliwa-Kopystynski, J., Burchell, M.J., Lowen, D., 2008. Impact cratering and break up of the small bodies of the Solar System. Icarus 195, 817–826. doi:10.1016/j.icarus.2008.02.010.

Melosh, H.J., 1984. Impact ejection, spallation, and the origin of meteorites. Icarus 59, 234–260. doi:10.1016/0019-1035(84)90026-5.

Melosh, H.J., 1989. Impact cratering. Oxford University Press, USA. 245pp.

Milner, D.J., Baldwin, E.C., Burchell, M.J., 2008. Laboratory investigations of marine impact events: Factors influencing crater formation and projectile survivability. Meteorit. Planet. Sci. 43, 2015–2026. doi:10.1111/j.1945-5100.2008.tb00658.x.

Nakamura, E. et al., 2012. Space environment of an asteroid preserved on micrograins returned by the Hayabusa spacecraft. Proc. Natl. Acad. Sci. USA 109, E624–E629.

Nakazawa, S., Watanabe, S.-I., Iijima, Y., Kato, M., 2002. Experimental investigation of shock wave attenuation in basalt. Icarus 156, 539–550. doi:10.1006/icar.2001.6729.

Oberbeck, V.R., Morrison, R.H., 1976. Candidate areas for in situ ancient lunar materials. Proc. Lunar Sci. Conf. 7th 2983–3005.

Poelchau, M.H., Kenkmann, T., Hoerth, T., Schäfer, F., Rudolf, M., Thoma, K., 2014. Impact cratering experiments into quartzite, sandstone and tuff: The effects of projectile size and target properties on spallation. Icarus 242, 211–224. doi:10.1016/j.icarus.2014.08.018.

Polanskey, C.A., Ahrens, T.J., 1990. Impact spallation experiments - Fracture patterns and spall velocities. Icarus 87, 140–155. doi:10.1016/0019-1035(90)90025-5.
20

Table 1.

| shot No. | Impact velocity (km/s) | Impact angle (deg.) | Target type | Target mass (kg) | Target curvature $\chi$ |
|---|---|---|---|---|---|
| 2900 | 3.447 | 90 | C9 | 0.8075 | 0.000 |
| 2901 | 3.454 | 90 | C9 | 0.7998 | 0.000 |
| 2902 | 3.381 | 90 | C9 | 0.8362 | 0.000 |
| 3019 | 3.431 | 90 | C9 | 0.8053 | 0.000 |
| 3020 | 3.150 | 90 | C9 | 0.8040 | 0.000 |
| 3018 | 3.580 | 90 | C15 | 3.0946 | 0.000 |
| 2702 | 3.404 | 82 | S25 | 4.2570 | 0.013 |
| 2926 | 3.425 | 82 | S25 | 2.6609 | 0.013 |
| 2927 | 3.554 | 78 | S25 | 2.2803 | 0.013 |
| 2703 | 3.381 | 78 | S17 | 1.4031 | 0.019 |
| 2921 | 3.227 | 77 | S17 | 1.4094 | 0.019 |
| 2922 | 3.430 | 79 | S17 | 1.4196 | 0.019 |
| 2930 | 3.401 | 83 | S17 | 1.3278 | 0.019 |
| 2699 | 3.415 | 88 | S11 | 0.7036 | 0.029 |
| 2920 | 3.120 | 74 | S11 | 0.7293 | 0.029 |
| 2931 | 3.437 | 67 | S11 | 0.7135 | 0.029 |
| 2704 | 3.694 | 72 | S8 | 0.2609 | 0.041 |
| 2904 | 3.526 | 70 | S8 | 0.2664 | 0.041 |
| 2906 | 3.388 | 71 | S8 | 0.2641 | 0.041 |
| 2907 | 3.636 | 69 | S8 | 0.2740 | 0.041 |

The projectile and target conditions of each shot. The impact angles were measured from the plane tangent to the impact point. $\chi$ is the normalized curvature defined as the ratio of the projectile diameter to the target diameter. The targets include the following six types: C9 and C15 are cubes with sides measuring 9 and 15 cm, respectively, S25 and S17 are hemispheres with diameters of 24.8 and 17.0 cm, respectively, and S11 and S8 are spheres with diameters of 10.9 and 7.8 cm, respectively.

| Table 2 | Whole crater | | | | Pit | | | Spall region | |
|---|---|---|---|---|---|---|---|---|---|
| | Volume | Area | Depth | Radius | Volume | Area | Radius | Volume | Area |
| shot No. | (mm³) | (mm²) | (mm) | (mm) | (mm³) | (mm²) | (mm) | (mm³) | (mm²) |
| 2900 | 675.6 | 241.8 | 8.42 | 8.8 | 505.7 | 79.0 | 5.0 | 169.9 | 162.8 |
| 2901 | 835.5 | 389.5 | 8.84 | 11.1 | 505.7 | 75.5 | 4.9 | 329.8 | 314.0 |
| 2902 | 642.6 | 260.8 | 8.41 | 9.1 | 474.7 | 74.2 | 4.9 | 167.9 | 186.6 |
| 3019 | 782.9 | 344.0 | 8.92 | 10.5 | 529.6 | 82.9 | 5.1 | 253.4 | 261.1 |
| 3020 | 707.8 | 345.8 | 8.40 | 10.5 | 444.4 | 70.4 | 4.7 | 263.4 | 275.4 |
| 3018 | 672.1 | 364.6 | 8.23 | 10.8 | 521.0 | 87.8 | 5.3 | 151.1 | 276.8 |
| 2702 | 764.4 | 320.4 | 8.68 | 10.1 | 506.5 | 77.4 | 5.0 | 257.9 | 243.0 |
| 2926 | 898.7 | 398.3 | 9.01 | 11.3 | 506.0 | 81.2 | 5.1 | 392.7 | 317.1 |
| 2927 | 884.6 | 379.4 | 8.82 | 11.0 | 553.6 | 83.5 | 5.2 | 331.0 | 295.9 |
| 2703 | 1167.1 | 476.7 | 8.86 | 12.3 | 497.1 | 74.6 | 4.9 | 670.0 | 402.1 |
| 2921 | 994.1 | 476.2 | 8.32 | 12.3 | 397.9 | 58.0 | 4.3 | 596.2 | 418.2 |
| 2922 | 1051.4 | 429.3 | 8.79 | 11.7 | 489.1 | 70.5 | 4.7 | 562.3 | 358.8 |
| 2930 | 1090.3 | 459.0 | 9.05 | 12.1 | 537.2 | 77.5 | 5.0 | 553.1 | 381.5 |
| 2699 | 1189.0 | 563.2 | 8.74 | 13.4 | 492.3 | 73.9 | 4.9 | 696.7 | 489.3 |
| 2920 | 1084.3 | 528.7 | 8.53 | 13.0 | 413.1 | 67.3 | 4.6 | 671.2 | 461.4 |
| 2931 | 981.8 | 422.6 | 8.59 | 11.6 | 470.2 | 73.2 | 4.8 | 511.6 | 349.4 |
| 2704 | 1637.8 | 697.3 | 8.72 | 14.9 | 608.7 | 88.1 | 5.3 | 1029.1 | 609.2 |
| 2904 | 1265.2 | 645.2 | 8.59 | 14.3 | 490.0 | 75.1 | 4.9 | 775.2 | 570.1 |
| 2906 | 1754.4 | 709.8 | 8.58 | 15.0 | 425.1 | 65.7 | 4.6 | 1329.3 | 644.1 |
| 2907 | 2826.4 | 938.3 | 8.96 | 17.3 | 502.3 | 71.3 | 4.8 | 2324.1 | 867.0 |

The dimensions of the resultant craters.

Figure 1

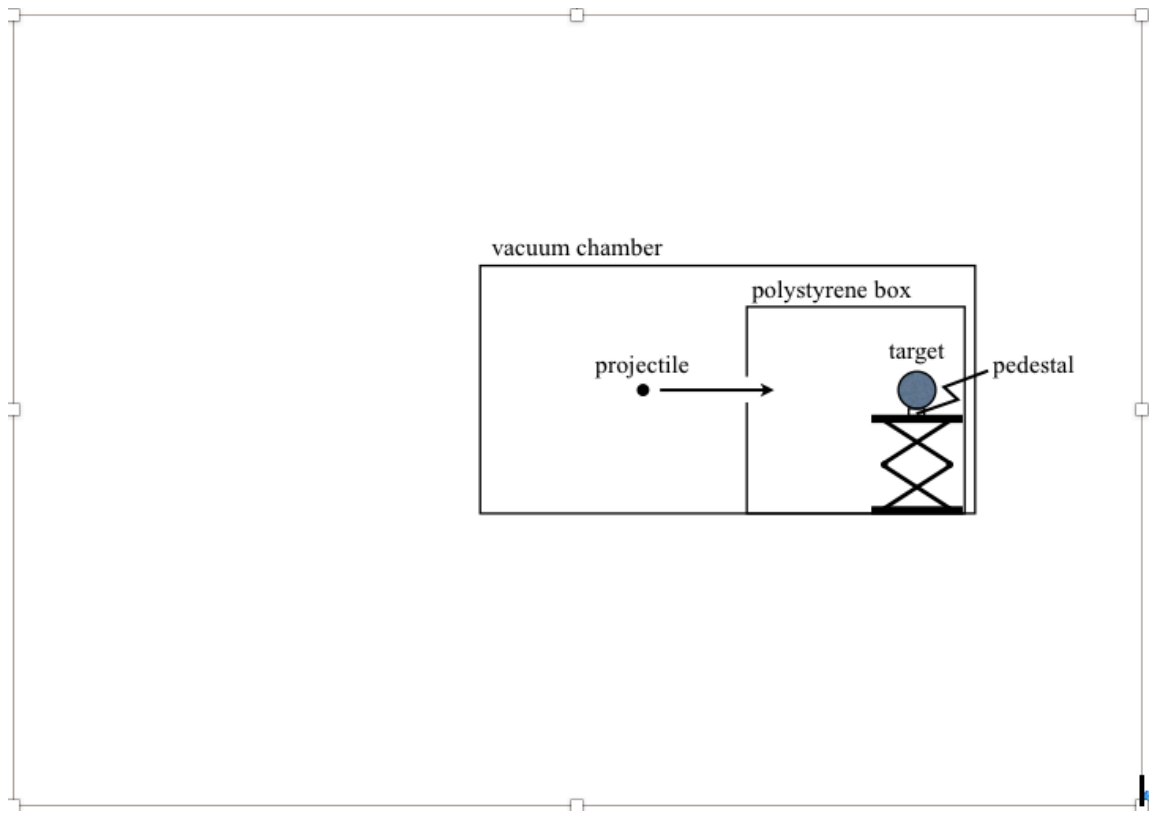

Schematic cross section of the experimental setup. The targets were placed on a mechanical jack with a pedestal, and this equipment was set in a polystyrene box with an acrylic window on the side for observations and a small rectangular hole at the front through which the projectile passed. The polystyrene box was placed in a vacuum chamber.

Figure 2

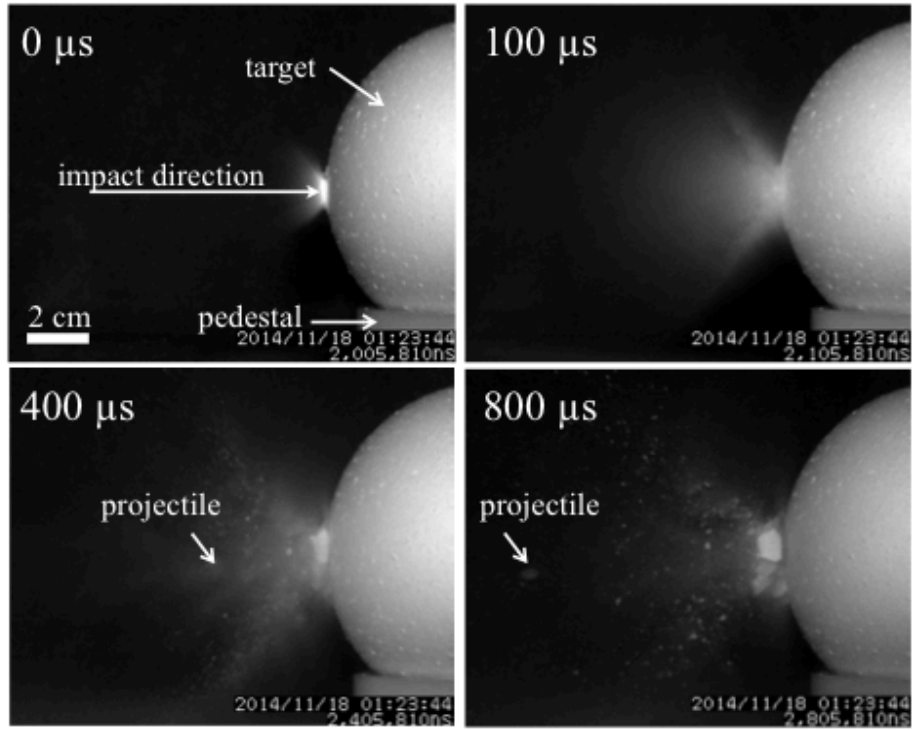

Snapshots of the impact on a spherical target (#2699) at the time of the impact (0 μs) and at 100, 400, and 800 μs after the impact. The projectile came from the left side of the figures.

Figure 3

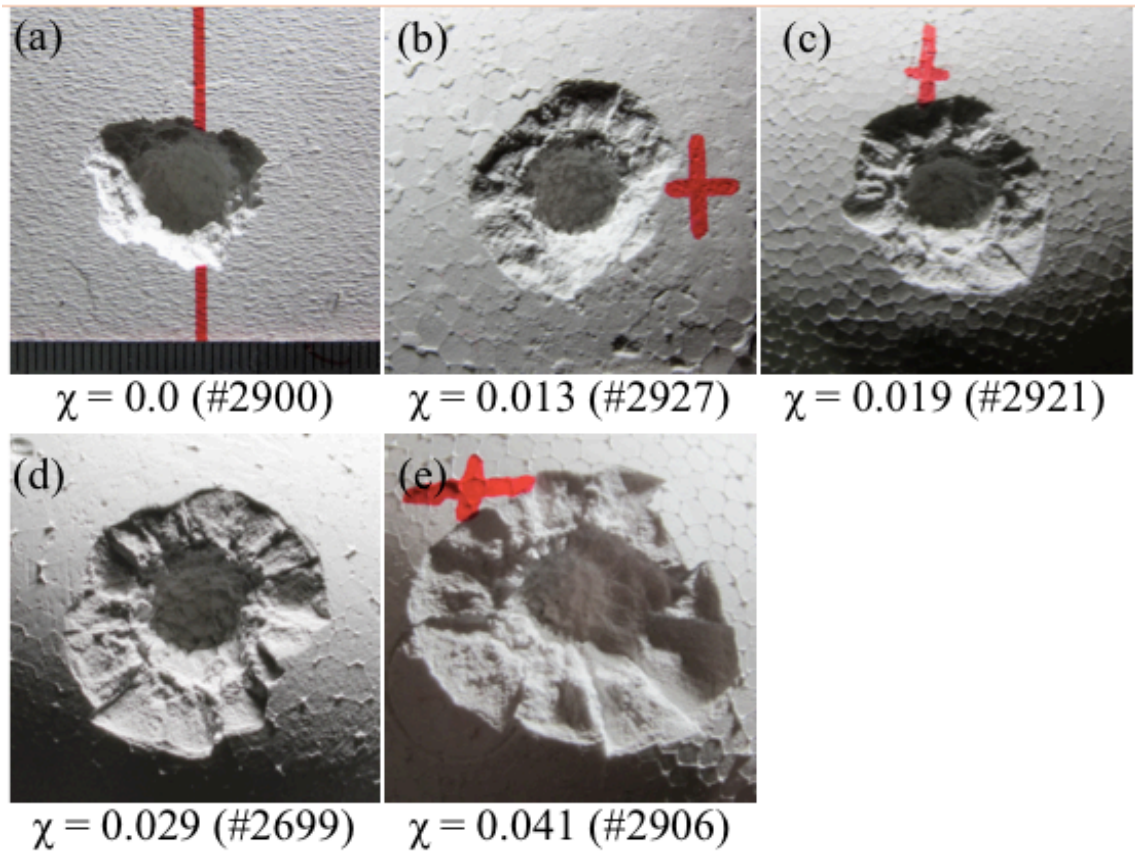

Photographs of typical resultant craters with different curvatures (on the same length scale). Each crater consists of a deep circular pit and an irregular-shaped spall region around the pit. The pit shows a rugged surface, while the surface of the spall region is smooth and seems to be the face of the fracture.

Figure 4

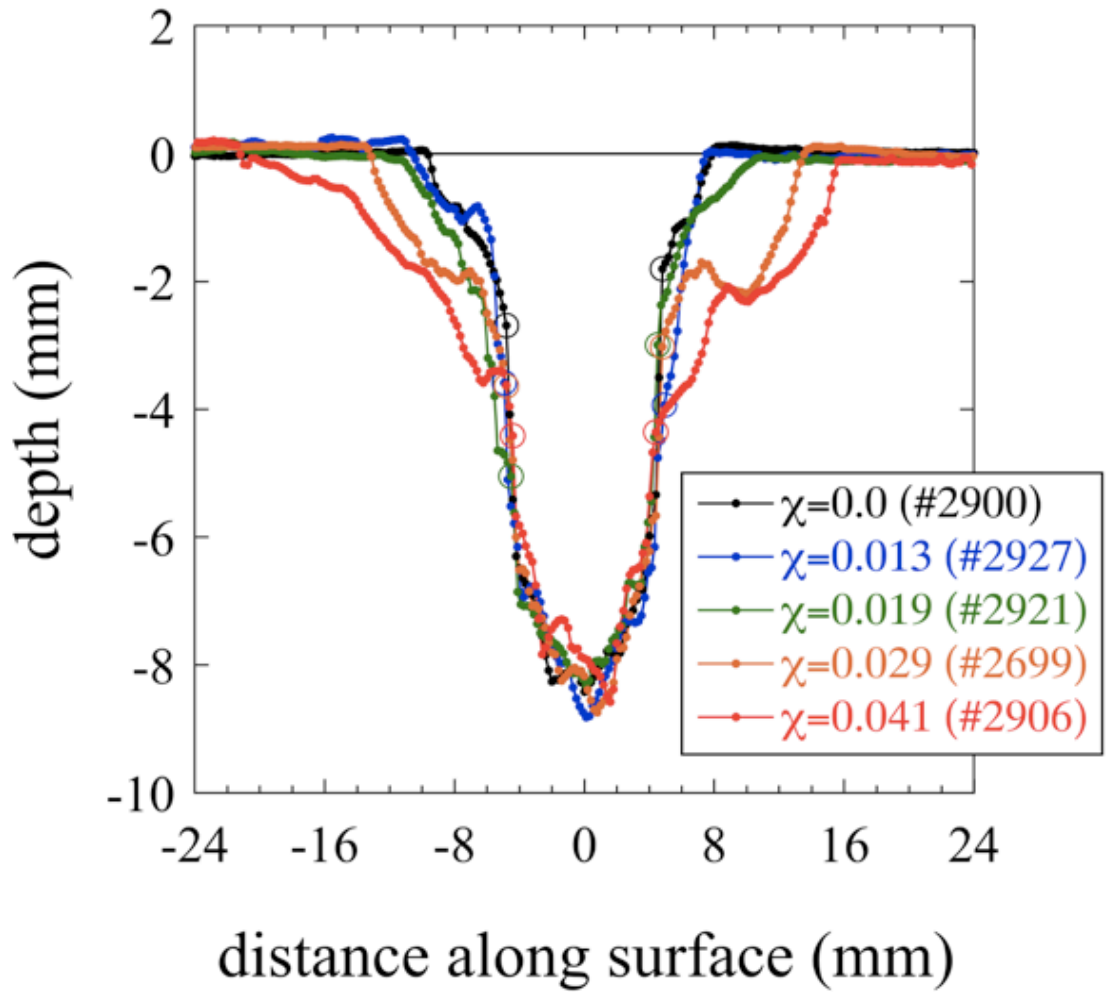

Cross-sectional topography of typical craters with various curvatures. The boundary between the pit and the spall region is marked with open circles.

Figure 5

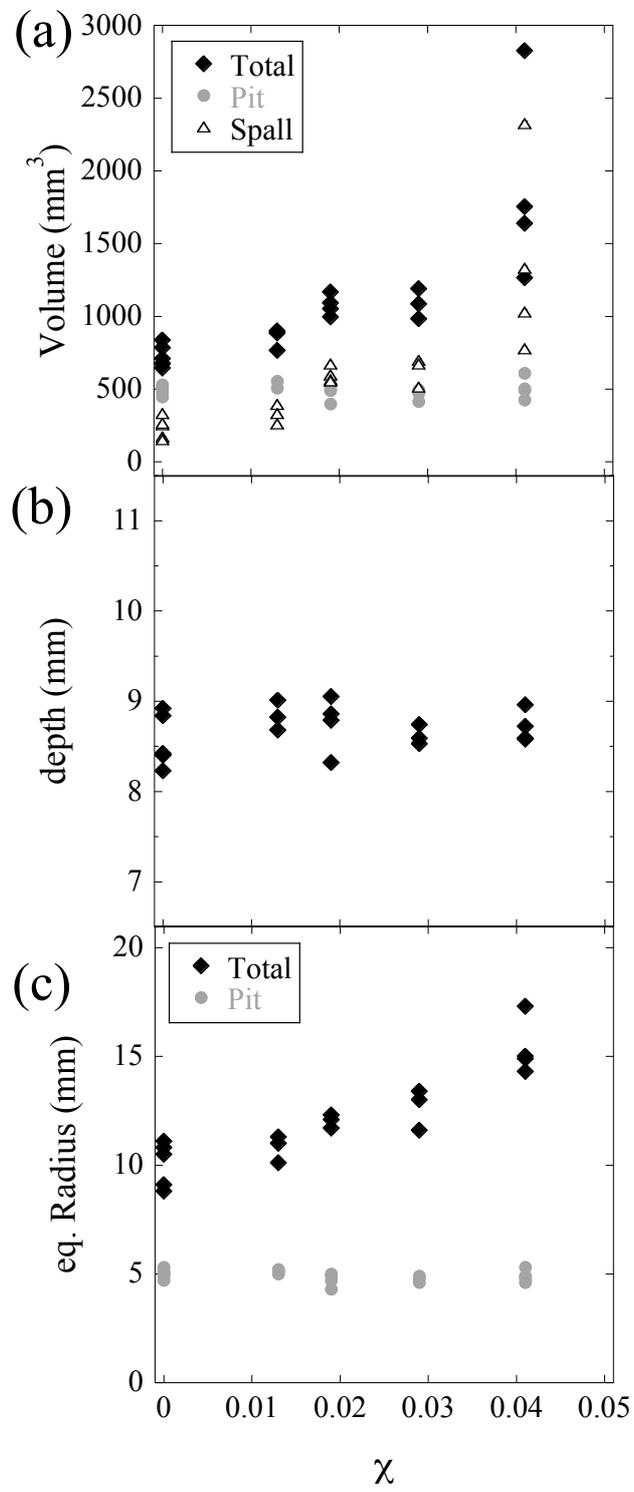

Crater dimensions as a function of normalized curvatures: (a) the volumes of whole craters, pits, and spall regions, (b) the depth of craters, and (c) the radius of craters.

Figure 6

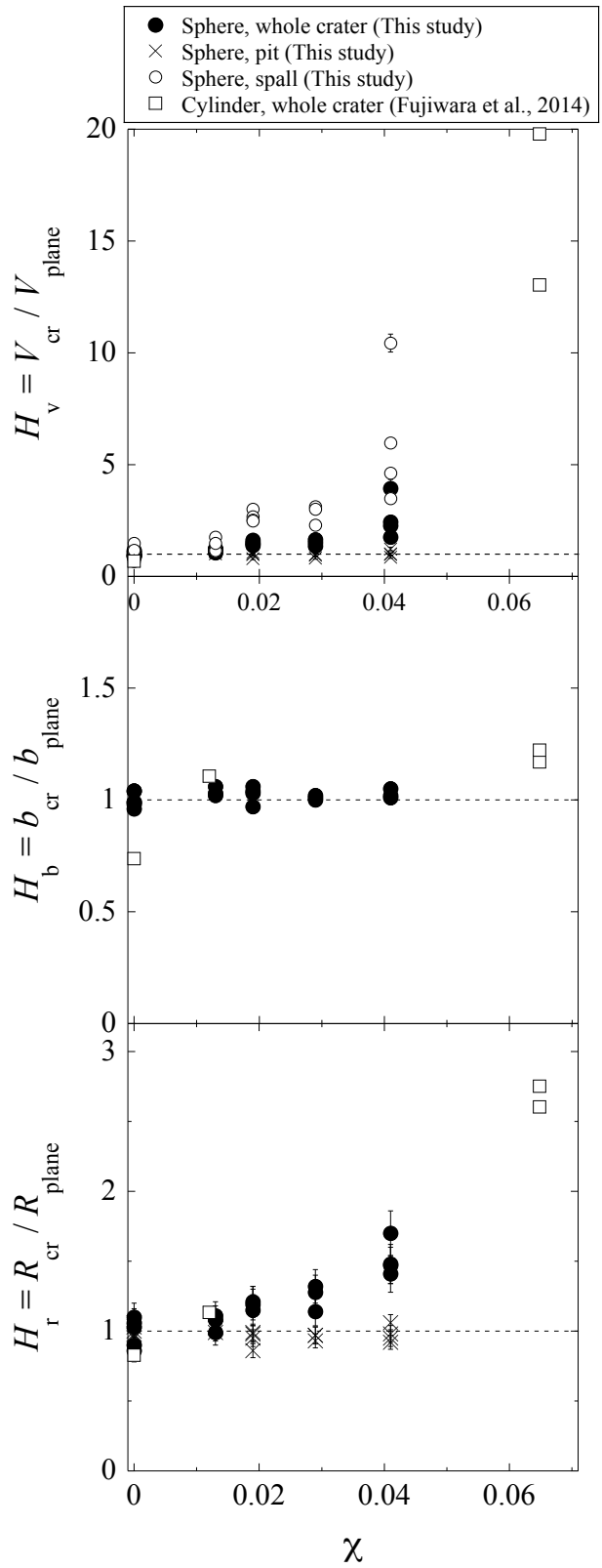

(a) The crater volume ($H_v$), (b) depth ($H_b$), and (c) radius ($H_r$) on curved targets normalized by those on plane targets are plotted as a function of the normalized curvature. $V_{cr}$, $b_{cr}$, $R_{cr}$ are the volume, depth, and radius of the craters on curved targets, and $V_{plane}$, $b_{plane}$, $R_{plane}$ are those on plane targets, respectively. We also plot the data derived from craters formed on the side of cylindrical gypsum by a nylon projectile that impacted the target at 3–4 km/s (Fujiwara et al., 2014).

Figure 7

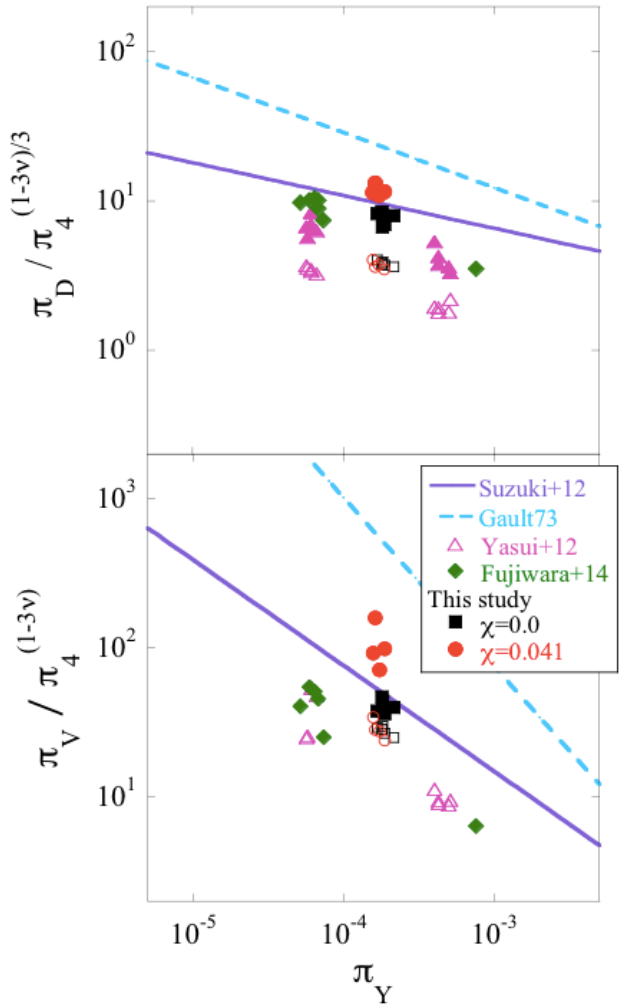

The scaled cratering efficiencies for diameter and volume calculated from our results are plotted against the dimensionless strength parameter. Filled symbols indicate the values for whole craters including spall regions, while open symbols indicate those for pits. Each error is within the size of the point. Solid and broken lines are the regression lines derived by cratering on sedimentary rocks (porosity ~15%; Suzuki et al., 2012) and on igneous rocks (porosity must be very low; Gault, 1973). We also plot the data obtained by impacts on gypsum targets (Yasui et al., 2012; Fujiwara et al., 2014). Suzuki+12,

Gault73, Yasui+12, and Fujiwara+14 denote Suzuki et al. (2012), Gault (1973), Yasui et al. (2012), and Fujiwara et al. (2014), respectively.

Figure 8

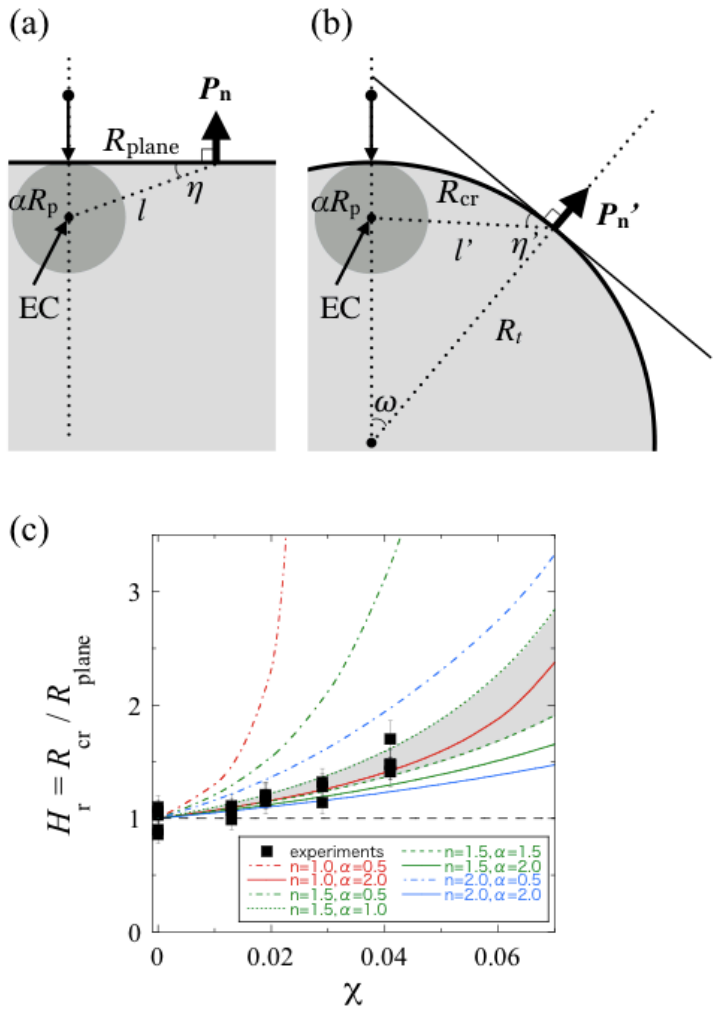

Schematic views of the impacts onto (a) planar and (b) curved ($\chi = R_p/R_t$) surfaces with variables used in our simple model. The projectile comes from the top along the vertical dotted line (indicated by the arrow with a black circle at the end). The dark-gray circles represent the isobaric core. (c) The model curves of various pairs of $n$ and $\alpha$ are plotted with the experimental results. The gray area indicates the area of $n = 1.5$ and $\alpha = 1.0$–1.5.

Figure 9

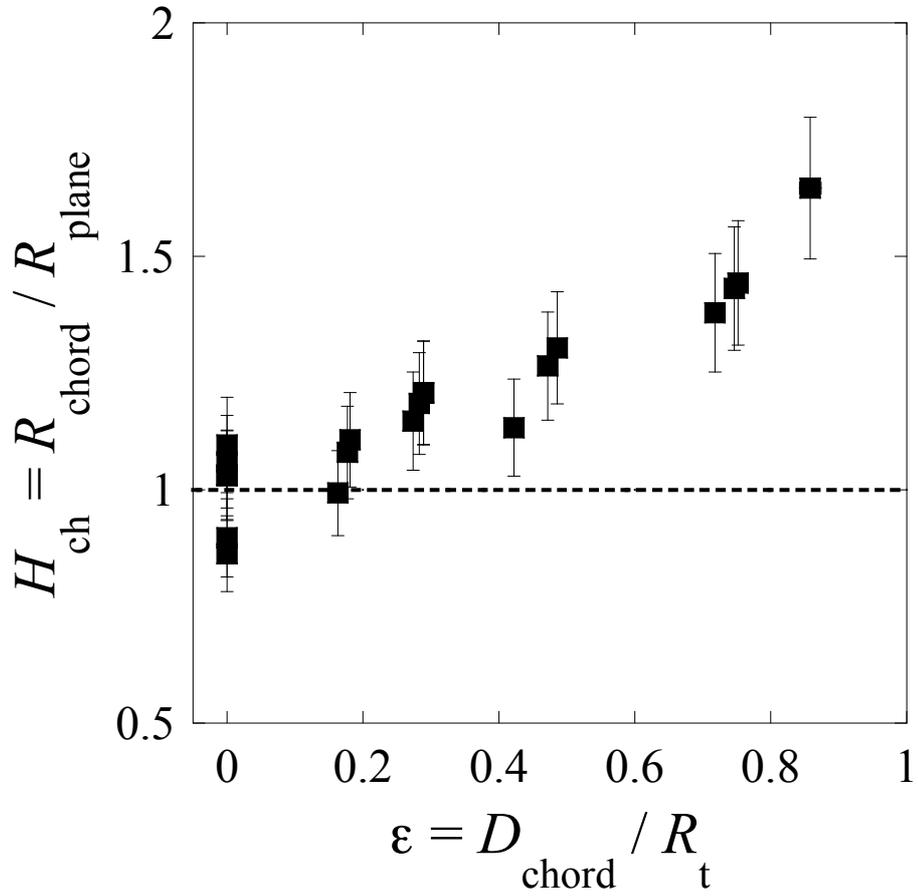

The experimental results are plotted on the diagram with parameters measurable for craters in the field.